\documentclass[prd,10pt,aps,showpacs,amsmath,amssymb,onecolumn]{revtex4-1}
\usepackage{graphicx}
\usepackage{dcolumn}
\usepackage{bm}
\usepackage{color} % allowing color
\usepackage[usenames,dvipsnames]{xcolor}
\usepackage{float}
\usepackage{verbatim}
\usepackage{soul}

 \def\be{\begin{equation}}
\def\ee{\end{equation}}

\usepackage{hyperref}

\begin{document}

\title{Very Magnetized White Dwarfs with Axisymmetric  Magnetic Field and the Importance of the Electron Capture and Pycnonuclear Fusion Reactions for their Stability}%Axisymmetric Magnetic Fields, Electron Capture and Pycnonuclear
  %Reactions in Magnetized White Dwarfs}

\author{Edson Otoniel$^{1,3}$} \email{edson.otoniel@gmail.com}

\author{Bruno Franzon$^{2}$} \email{franzon@fias.uni‐frankfurt.de}

\author{Manuel Malheiro$^{1}$} \email{malheiro@ita.br}

\author{Stefan Schramm$^{2}$} \email{schramm@fias.uni‐frankfurt.de}

\author{Fridolin Weber$^{3}$} \email{fweber@mail.sdsu.edu}

\affiliation{$^{1}$Departamento de F\'isica, Instituto Tecnol\'ogico
  de Aeron\'autica, Pra\c{c}a Marechal Eduardo Gomes, 50 - Vila das
  Acacias, 12228--900 S\~ao Jos\'e dos Campos, SP, Brazil}

\affiliation{$^{2}$Frankfurt Institute for Advanced Studies,
  Ruth-Moufang-1 60438 Frankfurt am Main, Germany}

\affiliation{$^{3}$Department of Physics, San Diego State University,
  5500 Campanile Drive, San Diego, California 92182, USA and \\ Center
  for Astrophysics and Space Sciences, University of California at San
  Diego, La Jolla, California 92093, USA}

\date{\today}

\begin{abstract}
In this work, we study the properties of magnetized white dwarfs
taking into account possible instabilities due to electron capture and
pycnonuclear fusion reactions in the cores of such objects. The
structure of white dwarfs is obtained by solving the Einstein-Maxwell
equations with a poloidal magnetic field in a fully general
relativistic approach. The stellar interior is composed of a regular
crystal lattice made of carbon ions immersed in a degenerate
relativistic electron gas. The onsets of electron capture reactions
and pycnonuclear reactions are determined with and without magnetic
fields.  We find that magnetized white dwarfs violate the standard
Chandrasekhar mass limit significantly, even when electron capture and
pycnonuclear fusion reactions are present in the stellar interior. We
obtain a maximum white dwarf mass of around $2.14\,M_{\odot}$ for a
central magnetic field of $\sim 3.85\times 10^{14}$~G, which indicates
that magnetized white dwarfs may play a role for the interpretation of
superluminous type Ia supernovae. Furthermore, we show that the
critical density for pycnonuclear fusion reactions limits the central
white dwarf density to $9.35\times 10^9$ g/cm$^3$. As a result, equatorial radii of white dwarfs cannot be smaller than $\sim 1100$~km.  Another interesting
feature concerns the relationship between the central stellar density
and the strength of the magnetic field at the core of a magnetized
white dwarf.  For high magnetic fields, we find that the central
density increases (stellar radius decrease) with magnetic field
strength, which makes ultramagnetized white dwarfs more compact.  The
opposite is the case, however, if the central magnetic field is less
than $\sim 10^{13}$~G. In the latter case, the central density
decreases (stellar radius increases) with central magnetic field
strengths.
\end{abstract}

\maketitle

\section{Introduction}

It is generally accepted that stars born with masses below around 10
solar masses end up their evolutions as white dwarfs (WDs)
\cite{weber1999pulsars,shapiro2008black,glendenning2012compact}. With
a typical composition mostly made of carbon, oxygen, or helium, white
dwarfs possess central densities up to $\sim 10^{11}\,{\rm
  g/cm}^{3}$. They can be very hot \citep{0004-637X-704-2-1605}, fast
rotating
\cite{arutyunyan1971rotating,boshkayev2013general,hartle1967slowly}
and strongly magnetized
\cite{2014ApJ...794...86C,lobato_magnetars_2016,Banibrata_2016}. The observed surface
magnetic fields range from $10^{6}\,$ G to $10^{9}\,$ G
\cite{Terada:2007br,Reimers:1995ia,Schmidt:1995eh,Kemp:1970zz,putney1995three,%
  angel1978magnetic}.
The internal magnetic fields of white dwarfs are not known, but they
are expected to be larger than their surface magnetic fields. This is
due to the fact that in ideal magneto hydrodynamics (MHD), the
magnetic field, $B$, is `frozen-in' with the fluid and $B\propto \rho$,
with $\rho$ being the local mass density (see, e.g.,
Refs.~\cite{mestel2012stellar,landau_quantum_1958}).  A simple
estimate of the internal magnetic field strength follows from the virial
theorem by equating the magnetic field energy with the gravitational
binding energy, which leads to an upper limit for the magnetic fields
inside WDs of about $\sim 10^{13}$ G.  On the other hand, analytic and
numeric calculations, both in Newtonian theory as well as in General
Relativity theory, show that WDs may have internal magnetic fields as
large as $10^{12-16}\,$ G (see,
e.g., Refs.~\cite{angel1978magnetic,shapiro2008black,das_maximum_2014,bera2014mass,bera2016mass,Franzon:2015gda,Franzon:2016gzf,das_grmhd_2015-1}).

The relationship between the gravitational stellar mass, $M$, and the
radius, $R$, of non-magnetized white dwarfs was first determined by
Chandrasekhar~\cite{chandrasekhar1939}. Recently, mass-radius
relationships of magnetic white dwarfs have been discussed in the
literature (see, e.g.,
Refs.~\cite{suh2000mass,bera2014mass,Franzon:2015gda}).  These studies
show that the masses of white dwarfs increase in the presence of
strong magnetic fields. This is due to the Lorentz force, which acts
against gravity, therefore supporting stars with higher masses.

Based on recent observations of several superluminous type Ia
supernovae (SN 2006gz, SN 2007if, SN 2009dc, SN 2003fg)
\cite{silverman2011fourteen, Scalzo:2010xd, Howell:2006vn,
  Hicken:2007ap,Yamanaka:2009dp, taubenberger2011high,Kepler11032007},
it has been suggested that the progenitor masses of such supernovae
significantly exceed the Chandrasekhar mass limit of $M_{\rm Ch}\sim
1.4\, M_{\odot}$ \citep{Ilkov11012012}. Super-heavy progenitors were
studied as a result of mergers of two massive white dwarfs
\cite{Moll:2013mpa,0004-637X-773-2-136,0004-637X-827-2-128}.
Alternatively, the authors of Ref.~\cite{liu2014one} obtained
super-Chandrasekhar white dwarfs for magnetically charged stars. In
addition, super-Chandrasekhar white dwarfs were investigated in the
presence of strong magnetic fields in
Refs.~\cite{das_revisiting_2014}.  In
Refs.~\cite{adam1986models,ostriker1968rapidly1}, WDs models with
magnetic fields were calculated in the framework of Newtonian
physics. A recent study of differential rotating, magnetized white
dwarfs has shown that differential rotation might increase the mass of
magnetized white dwarfs up to 3.1 $\rm{M_{\odot}}$
\cite{Subramanian21112015}.  Also, as shown in
Ref.~\cite{bera_massradius_2016}, purely toroidal magnetic field
components can increase the masses of white dwarfs up to $5\,
M_{\odot}$.

According to Refs.~\cite{das_strongly_2012}, effects of an extremely
large and uniform magnetic field on the equation of state (EOS) of a
white dwarf could increase its critical mass up to
$2.58 \, M_\odot$. This mass limit is reached for extremely
large magnetic fields of $\sim 10^{18}$ G. Nevertheless, as already
discussed in Refs.~\cite{coelho_dynamical_2014,chamel_stability_2013},
the breaking of spherical symmetry due to magnetic fields and
micro-physical effects, such as electron capture reactions and
pycnonuclear reactions, can severely limit the magnetic field inside
white dwarfs.

In Ref.~\cite{Franzon:2015gda}, mass-radius relationships of highly
magnetized white dwarfs were computed using a pure
degenerate electron Fermi gas. However, according to
Ref.~\cite{salpeter_energy_1961}, many-body corrections modify the EOS
and, therefore, the mass-radius relationship of white dwarfs.  The
purpose of our paper is two-fold. Firstly, we model white dwarfs using
a model for the equation of state which takes into account not only
the electron Fermi gas contribution, but also the contribution from
electron-ion interactions \cite{PhysRevD.92.023008}. Secondly, we
perform a stability analysis of the matter in the cores of white
dwarfs against electron capture and pycnonuclear fusion reactions.
The Landau energy levels of electrons are modified by relativistic
effects if the magnetic field strength is higher than the critical QED
magnetic field strength of $B_{cr}=4.4\times 10^{13}$~G. However, as
already shown in Ref.~\cite{Bera:2014wja}, the global properties of
white dwarfs, such as masses and the radii, are nearly independent of
Landau quantization. For this reason, we do not take into account magnetic fields effects in the equation of state to calculate the global properties of WD's. 

Our paper is organized as follows. In Sec.\ \ref{sec:stellarint}, we
discuss the stellar interior of white dwarfs and details of the
equation of state used in our study to model white dwarfs.  This is
followed, in Sec.\ \ref{sec:equations}, by a brief discussion of the
equations that are being solved numerically to obtain the structure of
stationary magnetized white dwarfs. In Sec.\ \ref{sec:axisymm}, we
briefly discuss the Einstein-Maxwell tensor and the metric tensor used
to solve Einstein's field equations of General Relativity. The results
of our study are discussed in Sec.\ \ref{sec:results}  and summarized
in Sec.\ \ref{sec:summary}.

\section{Stellar interior}\label{sec:stellarint}

The properties of fermionic matter have been studied many decades ago
in Refs.~\cite{salpeter_energy_1961,hamada_models_1961}.  Typically, a
white dwarf is composed of atomic nuclei immersed in a fully ionized
electron gas. In this work, we make use of the latest experimental
atomic mass data \cite{wang_ame2012_2012,audi_ame2012_2012} used to
determine the equation of state. Modifications of the equation of
state due to the interactions between electrons and atomic nuclei are
taken into account too. The model adopted to describe the nuclear
lattice was derived for the outer crust of a neutron star in
Refs.~\cite{pearson_properties_2011,chamel_stability_2013} and later
applied to WDs in Ref.~\cite{chamel_maximum_2014}. According to
Ref.~\cite{chamel_maximum_2014}, the cores of white dwarfs are
subjected to the degenerate electron and ionic lattice pressures. The
total pressure is then given by
\begin{equation}
 P = P_{e} + P_L(Z,Z') \, ,
\end{equation}
where $P_{e}$ denotes the electron pressure, determined in
\cite{salpeter_energy_1961}, and $P_L(Z,Z')$ is the lattice pressure
for two different type of ions. The lattice pressure is given by the
energy density of the ionic lattice (see
Ref.~\cite{pearson_properties_2011}),
\begin{equation}
P_L(Z,Z')=\frac{1}{3}\mathcal{E}_L \, ,
\end{equation}
with $Z$ and $Z'$ being the proton number of two different ions. In
our case, the white dwarf is composed of carbon ions, i.e.,
$Z'$=$Z$=12.  Following the Bohr-van Leeuwen theorem
\cite{pearson_properties_2011}, the lattice pressure of ions arranged
in a regular body-centered-cubic (bcc) crystal does not depend on the
magnetic field, apart from a small contribution due to the quantum
zero-point motion of ions. In this case, the lattice energy density
reads \cite{chamel_stability_2013}
\begin{equation}
 \mathcal{E}_L=Ce^2n_e^{4/3}G(Z,Z') \, ,
 \label{sec2:energy}
\end{equation}
with $G(Z,Z')$ given by
\begin{equation}
 G(Z,Z')=\frac{\alpha Z^2+\gamma Z'^2+(1-\alpha-\gamma)ZZ'}{(\xi Z
   +(1-\xi)Z')^{4/3}} \, .
 \label{sec2:G}
\end{equation}
\begin{figure}[t!]
\centering
\vspace{0.1cm}
\includegraphics[width=1.5\textwidth,angle=0,scale=0.48]{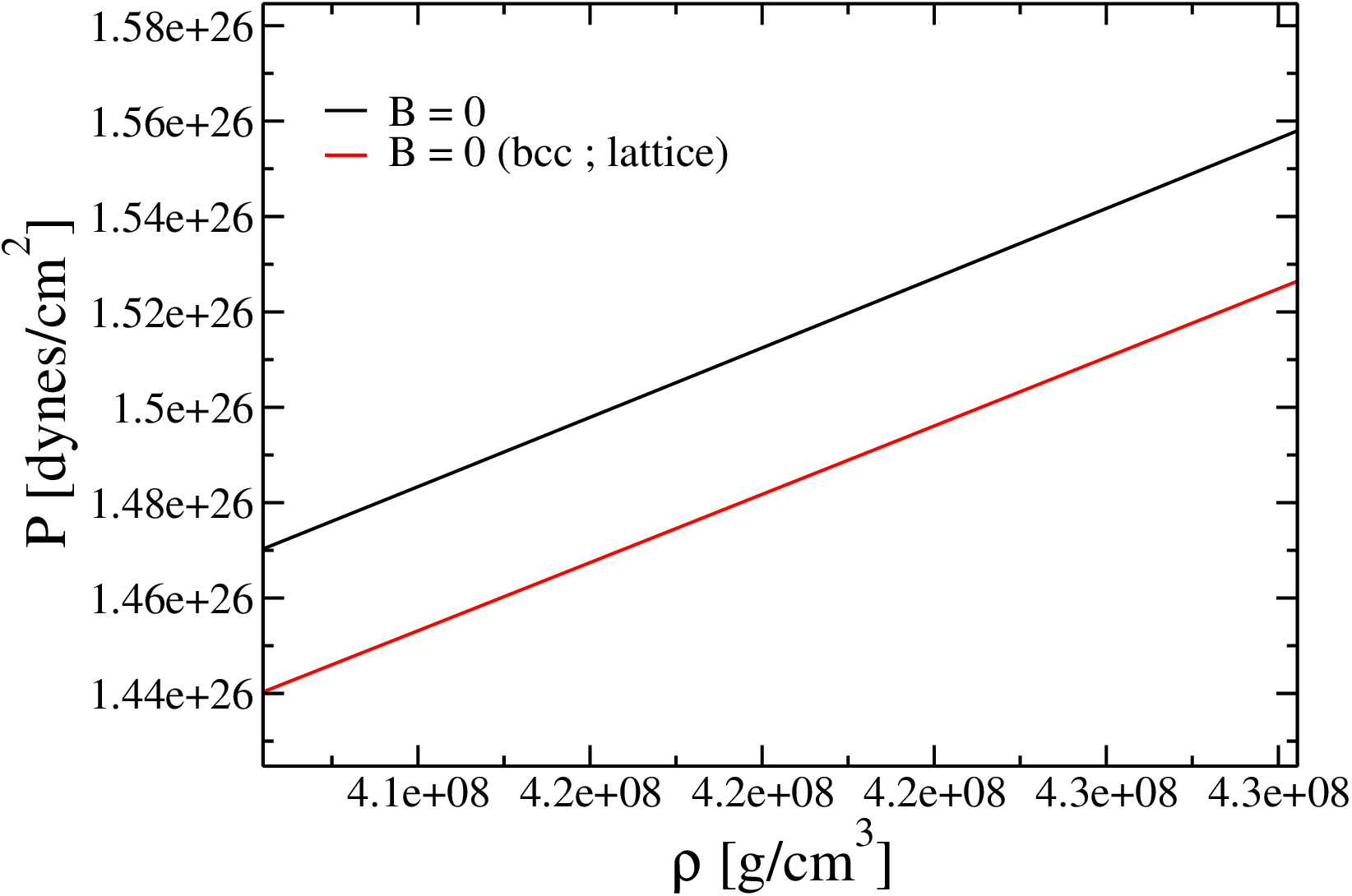}
\caption{(Color online) Equation of state for $B=0$ with (red
  curve) and without (black curve) lattice contributions.} 
\label{eos_carbono}
\end{figure}
The quantities $C$, $\alpha$, $\gamma$ are lattice constants and $\xi$
is the ratio of ions $^{A}_{Z}Y$ and $^{A'}_{Z'}Y$ in the lattice
\cite{chamel_maximum_2014} (see also Table \ref{tab:lattice
  Constant}). If only a single ion is present in the lattice,
Eq.~\ref{sec2:G} does not depend on $\alpha$ and $\gamma$ so that
Eq.~(\ref{sec2:energy}) becomes
\begin{equation}
 \mathcal{E}_L=Ce^2n_e^{4/3}Z^{2/3} \, ,
 \label{sec3:energy}
\end{equation}

The energy density is given in terms of the degenerate electron
energy, the energy density of the ions, and the energy density of the
ionic lattice,
\begin{equation}
\mathcal{E} = n_xM(Z,A) c^2 + n_{x'} M(Z',A') c^2 + \mathcal{E}_e +
\mathcal{E}_L-n_e m_e c^2 \, ,
\end{equation}
where $n_x$ and $n_x^{\prime}$ are the number densities of atomic
nuclei with masses $M(Z,A)$ and $M(Z',A')$, respectively. As already
mentioned above, here we adopt the most recent experimental values for
$M$ (see Refs.~\cite{wang_ame2012_2012,audi_ame2012_2012}).

Figure~\ref{eos_carbono} shows the impact of lattice contributions on
the white dwarf equation of state studied in this paper.  The black
lines (no lattice contribution) and red dashed line (with lattice
contribution) are for white dwarf matter with zero magnetic field
($B=0$). One sees that adding the lattice contribution to the equation
of state lowers the pressure somewhat, which in turn makes white
dwarfs less massive.  It also follows from this figure that the
presence of lattice contributions reduces the radii of white dwarfs
(renders them more compact) with comparable central pressures.
\begin{table}[tbp]
  \centering
  \caption{\label{tab:lattice Constant} Lattice constants $C$, $\alpha$,
  $\gamma$ and parameters $(1-\alpha-\gamma)$ and $\xi$ for a
  body-centered-cubic (bcc) structure, as obtained by the method of
  Coldwell-Horsfall and Maradudin (see
  Ref.~\cite{chamel_maximum_2014}) for more details.}
\begin{tabular}{cccccc}
\hline
\hline
Lattice & \textbf{$C$} & \textbf{$\alpha$} &\textbf{$\gamma$}&
\textbf{$(1-\alpha-\gamma)$}&\textbf{$\xi$}\\
\hline
bcc & -1.444231 & 0.389821 & 0.389821 &    0.220358&0.5 \\
\hline
\end{tabular}
\end{table}

\section{Instabilities in strongly magnetized white dwarfs}\label{sec:equations}

\subsection{Inverse $\beta$ decay}

As shown in Refs.~\cite{gamow_physical_1939,chamel_maximum_2014}, the
matter inside of white dwarfs is unstable due to inverse
$\beta$-decay,
\[ A(N,Z) + e^- \rightarrow A(N+1, Z-1) + \nu_e \, . \]
Because of this reaction, atomic nuclei become more neutron rich and
the energy density of the matter is being reduced, at a given
pressure, leading to a softer EoS. Using the thermodynamic relation
(at zero temperature) $\mathcal{E}_e + P_e = n_e\mu_e$, one obtains
the Gibbs free energy, $g$, per nucleon as
\begin{equation}
g(Z,z') =mc^2+\frac{\xi}{\xi
  A+(1-\xi)A'}\Delta(A,Z)+\frac{(1-\xi)}{\xi A+(1-\xi)A'}\Delta(A',Z')
+ \gamma_e\left[\mu_e+m_ec^2+\frac{4}{3}\frac{\mathcal{E}_L}{n_e}
  \right] \, ,
\label{secA:eq1}
\end{equation}
with $m$ being the neutron mass and $\Delta(A, Z)$ denoting the excess
mass of nuclei, which, for magnetic field strengths $<10^{17}$~G, is
independent of the magnetic field, see, e.g.,
Ref~\cite{chamel_stability_2013}. For $\gamma_e={\bar Z}/A$ we have
$\bar{Z}=\xi Z+(1-\xi)Z'$ and $\bar{A}=\xi A+(1-\xi)A'$, with $\mu_e$
being the electron chemical potential. Inverse $\beta$-decay reactions
are believed to occur in the cores of white dwarfs if the condition
\cite{chamel_maximum_2014}
\begin{equation}
 g(Z,Z')\geq g(Z-\Delta Z,Z'-\Delta Z')
 \label{secA:gibb}
\end{equation}
is fulfilled, where $g(Z,Z')$ and $g(Z-\Delta Z,Z'-\Delta Z')$ follow from
Eq.\ (\ref{secA:eq1}) and the possible choices for $\Delta Z$ and $\Delta Z'$ 
are $\Delta Z=1 ~\&~ \Delta Z'=0$, $\Delta Z=0~ \& ~\Delta Z'=1$, and $\Delta Z'=1 ~\&
~\Delta Z' =1$.

From the inequality (\ref{secA:gibb}), we obtain the following relation
\begin{equation}
\Delta\bar{Z}\left[\mu_e+\frac{4}{3}Ce^2n_e^{1/3}
  \Delta(\bar{Z}G(Z,Z'))\right]\geq\bar{\mu}_e^{\beta}
 \label{sec4:ineq2}
\end{equation}
with the electron number density $n_e$ and mass density $\rho$ of a
magnetized electron gas given respectively by
\begin{eqnarray}
 n_e &=& \frac{2B_{\star}}{(2\pi)^2\lambda^3}\sum_\nu g_{\nu 0}
 \sqrt{x_F^2-1- 2 \nu B_{\star}}. \\ \label{n_e}
 \rho&=&\frac{1}{\gamma_e} m n_e \, .
\end{eqnarray}
where only the ground-state Landau level $\nu=0$ is occupied,
$\nu_{\rm max}=1$. For two occupied levels, $\nu=0$ and $\nu=1$, one has
$\nu_{\rm max}=2$, and similarly for the higher levels. The quantities
$x_F$ in Eq.\ (\ref{n_e}) and $\bar\mu_e^{\beta}$ in Eq.\ (\ref{sec4:ineq2})
are defined as $x_F \equiv p_F/m_ec$ and 
\begin{eqnarray}
 \bar{\mu}_e^{\beta}=\xi \mu_e^{\beta}(A,Z)+(1-\xi)
 \mu_e^{\beta}(A',Z') \, ,
\end{eqnarray}
with $\mu_e^{\beta}(A,Z)$ and $\mu_e^{\beta}(A',Z')$ given by
\begin{eqnarray}
 \mu_e^\beta(A,Z)&\equiv& \Delta(A,Z-\Delta Z)-\Delta(A,Z)+m_ec^2 \\
 \mu_e^\beta(A',Z')&\equiv& \Delta(A',Z'-\Delta Z')-\Delta(A',Z')+m_ec^2  \, .
\end{eqnarray}
Another important quantity is $\Delta \bar{Z}G(Z,Z')$, which describes
the difference of $G$, defined in Eq.\ (\ref{sec2:G}), before and
after an inverse $\beta$ decay reaction. It is given by
\begin{equation}
 \Delta(\bar{Z}G(Z,Z'))=G(Z,Z')-G(Z-\Delta,Z'-\Delta) \, .
\end{equation}
For an electron gas consisting of only one type of ion, we have
\begin{equation}
\Delta(\bar{Z}G(Z,Z))=Z^{5/3}-(Z-1)^{5/3}- \frac{2}{3}Z^{2/3} \, .
\end{equation}

In the limit where only the ground state ($\nu=0$) is fully occupied
by electrons, one has
\[n_e=n_{eB}\propto B_{\star}^{2/3} \, ,\]
where $B_{\star} = B/B_c$ with $B_c=4.414\times10^{13}$~G being the
critical magnetic field (see Ref.\ \cite{haensel2007neutron} for more
details about $n_{eB}$). The chemical potential of the electrons in
this case is given by
\begin{equation}
\mu_e\approx \frac{2\pi^2m_ec^2\lambda_e^3n_{eB}}{B_{\star}} \, ,
  \label{sec4:mu}
\end{equation}
where $\lambda_{e}= \hbar/m_ec$ denotes the Compton wavelength of
electrons.  In Ref.\ \cite{chamel_maximum_2014} it was estimated that
the maximum magnetic field inside of white dwarfs, before the onset of
$\beta$-inverse reactions, is given by
\begin{equation}
B_{\star}^{\beta}\approx\frac{1}{2}
\left(\frac{\bar{\mu}_e^{\beta}(A,Z)}{m_ec^2
  \Delta\bar{Z}}\right)^2 \left[ 1 + \left( \frac{4}{\pi}\right)^{2/3}
  \frac{C\alpha}{3}\Delta(\bar{Z}G(Z,Z')) \right]^{-2} \, ,
\label{sec4:B}
\end{equation}
with $\alpha = e^2/(\hbar c)$ the fine structure constant. We note
that because of the second term on the right-hand-side of
Eq.\ (\ref{sec4:B}), which originates from lattice contributions, the
maximum value of $B_{\star}^{\beta}$ increases if lattice
contributions are taking into account.

\begin{figure}[t!]
\centering
\includegraphics[width=1.5\textwidth,angle=0,scale=0.48]{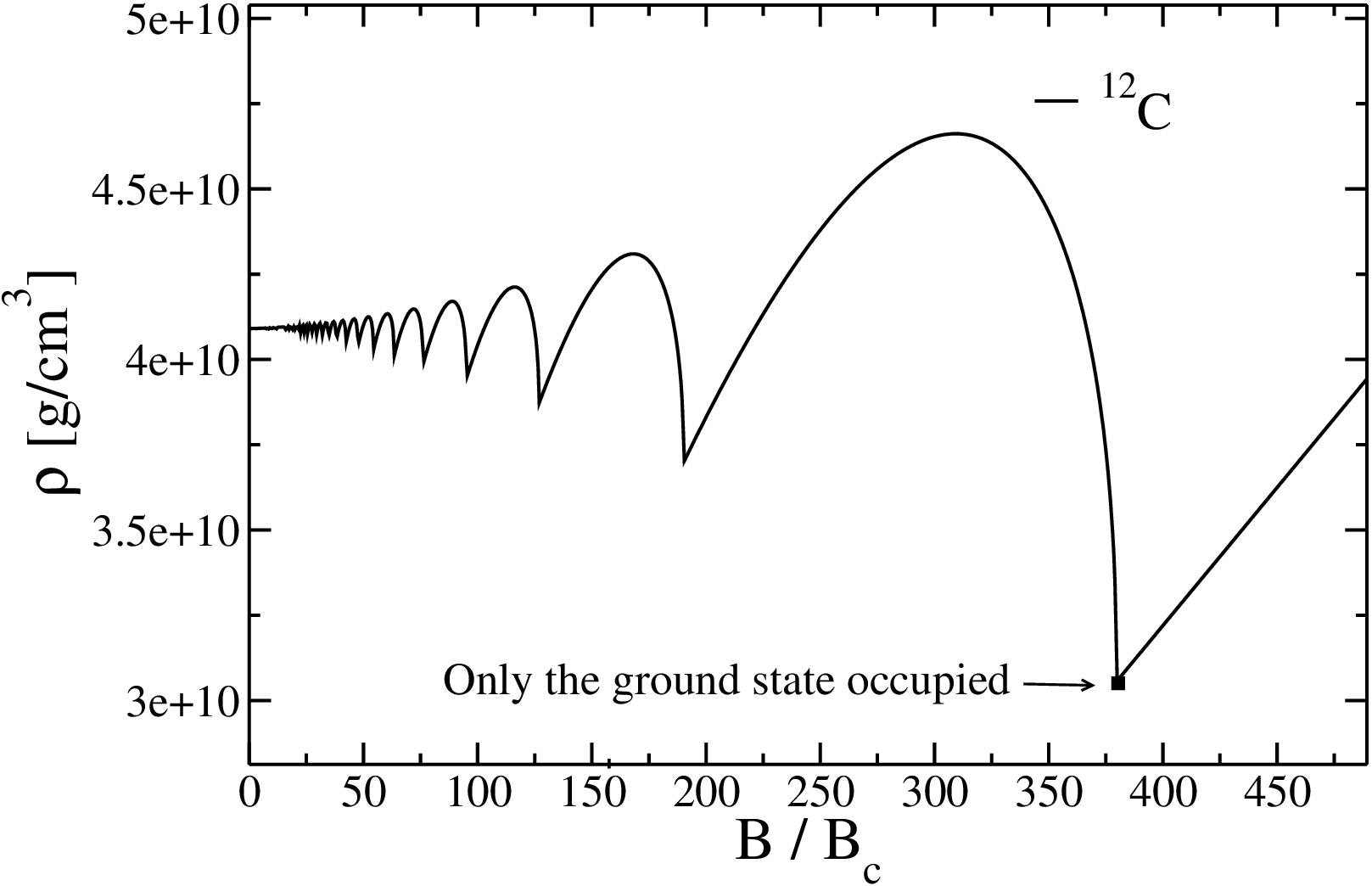}
\caption{Mass density thresholds for the onset of electron capture as
  a function of magnetic field strength (in units of the critical
  magnetic field, $B_{c}$), computed from Eq.\ (\ref{sec4:ineq2}) for
  matter made of only carbon ions.}
\label{rho_den}
\end{figure}

In Fig.~\ref{rho_den} we show the numerical solution of
Eq.\ (\ref{sec4:ineq2}) for white dwarf matter made of only carbon
ions immersed in a magnetized electron gas. The oscillatory behavior
is caused by the Landau level contributions to the number density,
given by Eq.\ (\ref{n_e}). For high values of $B$ with only the ground
state occupied, the dependence of density on $B$ becomes linear, as
can seen in Fig.\ \ref{rho_den}.

\subsection{Pycnonuclear reaction}
  
In this section, we will focus on nuclear fusion reactions
(pycnonuclear fusion reactions) among heavy atomic nuclei
,$^{A}_{Z}Y$, schematically expressed as $^{A}_{Z}Y + ^{A}_{Z}Y
\rightarrow ^{2A}_{2Z}Y$.  An example of such a reaction is carbon on
carbon, $^{12}\textrm{C}+^{12}\textrm{C}$.  Pycnonuclear reactions
have been found to occur over a significant range of stellar densities
(see, for instance, Ref.~\cite{gasques_nuclear_2005}), including the
density range found in the interiors of white dwarfs
\cite{chamel_stability_2013, chamel_maximum_2014}. The nuclear fusion
rates at which pycnonuclear reactions proceed, however, are highly
uncertain because of some poorly constrained parameters (see
Ref.\ \cite{gasques_nuclear_2005,yakovlev_fusion_2006}). The reaction
rates have been calculated for different models. In
Ref.\ \cite{gasques_nuclear_2005}, the pycnonuclear reaction rates are
defined as
\begin{equation}
R_{\textrm{pyc}}=\frac{n_i}{2}S(E_{\textrm{pk}}) \frac{\hbar}{mZ^2e^2} \, 
P_{\textrm{pyc}} \, F_{\textrm{pyc}}
\label{eq:Rpyc}
\end{equation}
where $S(E_{\textrm{pk}})$ is the astrophysical S-factor used in
Ref.~\cite{gasques_nuclear_2005} for the NL2 nuclear model
parametrization. According to Ref.~\cite{gasques_nuclear_2005}, an
analytic equation for the S-factor is given by
\begin{equation}
S(E_{\textrm{pk}}) = 5.15\times10^{16} \exp \left[ -0.428E_{\textrm{pk}}
  - \frac{3E_{\textrm{pk}}^{0.308}} {1+e^{0.613(8-E_{\textrm{pk}})}}
  \right] \, ,
\label{sec4:s}
\end{equation}
where $S(E_{\textrm{pk}})$ is in units of MeV barn. The factors
$P_{\textrm{pyc}}$ and $F_{\textrm{pyc}}$ in Eq.\ (\ref{eq:Rpyc})
are given by 
\begin{eqnarray}
  P_{\textrm{pyc}} &=& \exp\Big(-C_{\textrm{exp}}/\sqrt{\lambda}\Big) \, ,
  \label{eq:invl}
\\ F_{\textrm{pyc}} &=&
8C_{\textrm{pyc}}11.515/\lambda^{C_{\textrm{pl}}} \, ,
\label{eq:invl_2}
\end{eqnarray}
with $C_{\textrm{exp}}$, $C_{\textrm{pyc}}$ and $C_{\textrm{pl}}$ are
dimensionless parameters for a regular bcc-type crystal lattice (see
at zero temperature). Their values are listed in Table
\ref{tab:pyc_const}.
\begin{table}[h]
  \centering
  \caption{\label{tab:pyc_const} Coefficients $C_{\textrm{exp}}$,
    $C_{\textrm{pyc}}$, $C_{\textrm{pl}}$ related to pycnonuclear
    reaction rates at zero temperature, computed for nuclear model NL2
    (see Refs. \cite{PhysRevLett.78.3270,PhysRevLett.79.5218}).}
\begin{tabular}{cccc}
\hline
\hline
Model &$C_{\textrm{exp}}$ & $C_{\textrm{pyc}}$ &$C_{\textrm{pl}}$\\
\hline
bcc; static lattice & 2.638 & 3.90 & 1.25  \\
\hline
\end{tabular}
\end{table}

The inverse-length parameter $\lambda$ in Eq.\ (\ref{eq:invl}) and Eq.\ (\ref{eq:invl_2}) has the
form Refs.\ \cite{gasques_nuclear_2005,yakovlev_fusion_2006}
%\ot{NOTE: should we use a different symbol for $\lambda$ to avoid
%  confusion with the Compton wavelength?}
\begin{equation}
\lambda =\frac{\hbar^2}{mZ^2e^2}\Big(\frac{n_i}{2}\Big)^{1/3} =
\frac{1}{AZ^2} \left( \frac{1}{A}\frac{\rho X_i}{1.3574 \times
  10^{11}\textrm{g~cm}^{-1}} \right)^{1/3} \, .
\end{equation}

For number densities $\rho$ less than  neutron drip density one has $X_i=1$
\cite{gasques_nuclear_2005} and the pycnonuclear reaction rates are given by
\begin{equation}
R_{\textrm{pyc}} = \rho X_iAZ^4S(E_{\textrm{pk}}) C_{\textrm{pyc}}
10^{46} \lambda^{3-C_{\textrm{pl}}} \exp\Big( - C_{\textrm{exp}} /
\sqrt{\lambda} \Big) \, ,
\end{equation}
with $R_{\textrm{pyc}}$ given in units of cm$^{-3}$~s$^{-1}$.  The
zero-point oscillation energy $E_{\textrm{pk}}$ of $^{12}$C nuclei at
$\rho=10^{10}$~g/cm$^3$ is given by Ref.\ \cite{shapiro2008black}
\begin{equation}
E_{\rm pk}=\hbar\omega=\hbar \left( \frac{4\pi e^2Z^2\rho}{A^2M^2}
\right)^{1/2} \, .
\end{equation}
\begin{figure}[t!]
\centering
\includegraphics[width=0.49\linewidth,angle=0]{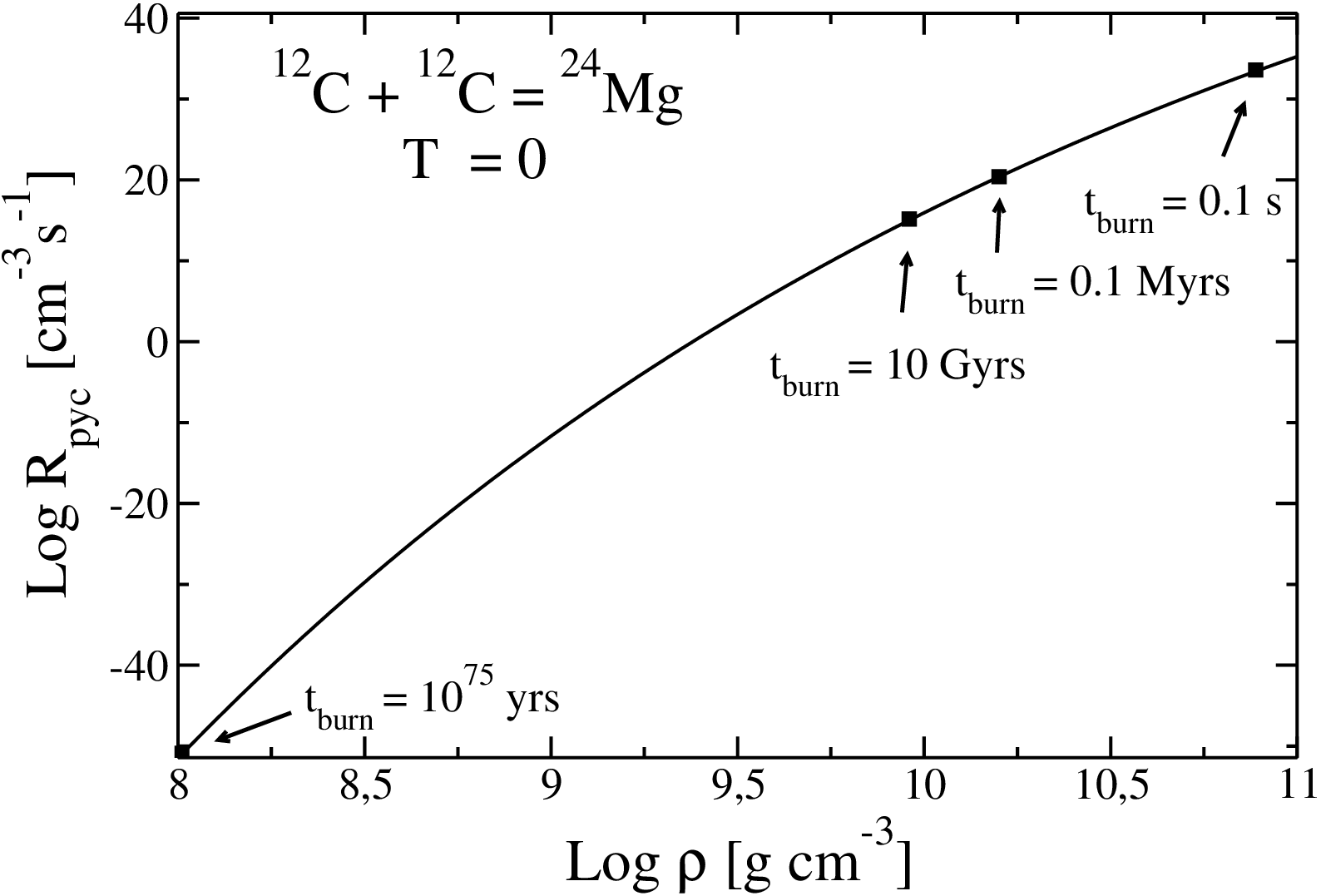}
\hfill
\includegraphics[width=0.48\linewidth,angle=0]{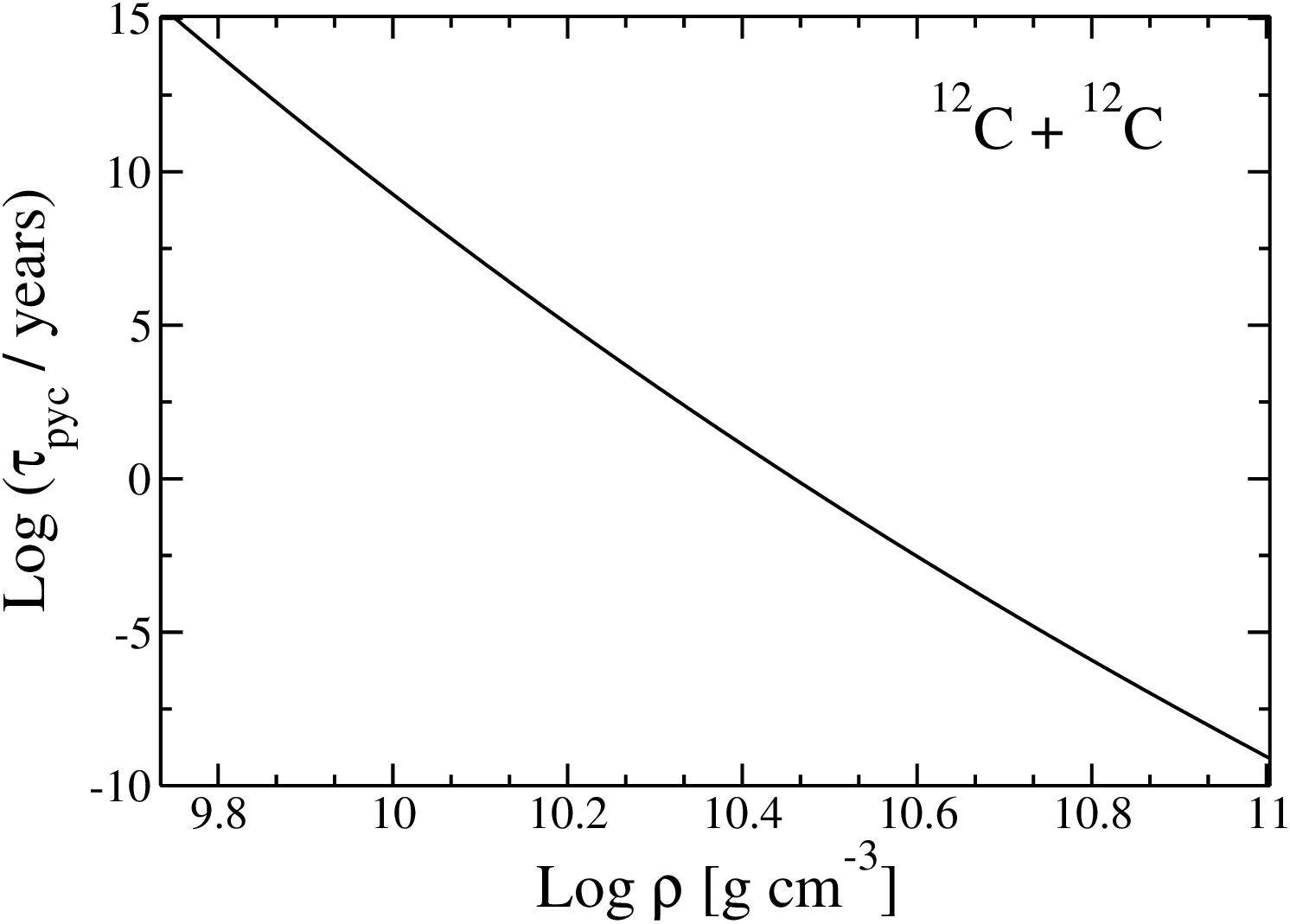}
\caption{\label{RhoxM_lam} Left: pycnonuclear fusion reaction rates
  for carbon burning at zero temperature as functions of mass density,
  for nuclear model NL2 and a bcc crystal lattice. Right: pycnonuclear
  reaction time scales at zero temperature for C+C fusion as a
  function of mass density. The S-factor is given by
  Eq.\ (\ref{sec4:s}) and the zero-point oscillation energy is $E_{pk}
  \sim 0.034$~MeV.}
\end{figure}

The time it takes for the complete fusion of atomic nuclei of mass
$Am$ is obtained from
\cite{,boshkayev2013general,gasques_nuclear_2005}
\begin{equation}
\tau_{\textrm{pyc}} = \frac{n_x} {R_{\textrm{pyc}}} = \frac{\rho}
    { {\rm Am} R_{\textrm{pyc}}} \, .
\end{equation}
As already mentioned above, the reaction rates are rather uncertain,
and the analytic astrophysical S-factor has an uncertainly of $\sim$
3.5, which considerably affects the density thresholds of pycnonuclear
reaction and their reaction times. Finally, in Fig.~\ref{RhoxM_lam} we show pycnonuclear fusion reaction rates and pycnonuclear reaction time scales 
  for carbon burning at zero temperature as functions of mass density. The  bcc crystal lattice for nuclear model NL2 was employed to produce Fig.~\ref{RhoxM_lam}.

\section{White dwarfs with axisymmetric magnetic fields}\label{sec:axisymm}

The numerical technique used in this work to study axisymmetric
magnetic fields was first applied to neutron stars in
Refs.~\cite{Bonazzola:1993zz, Bocquet:1995je}, and more recently in
Ref.~\cite{franzon2016self, Franzon:2016iai, Franzon:2016urz}. The
same formalism was used to study rotating and magnetized white dwarfs
in Ref.~\cite{Franzon:2015gda}. Here we build stellar equilibrium
configurations by solving the Einstein-Maxwell field equations in a
fully general relativistic approach. For more details about the
theoretical formalism and numerical procedure, see, for instance,
Ref.~\cite{gourgoulhon20123+}.  Below we show the basic
electromagnetic equations which, combined with the gravitational
equations, are solved numerically by means of a spectral method. In
this context, the stress-energy tensor $T_{\alpha\beta}$ is composed
of the matter and the electromagnetic source terms, \be
T_{\alpha\beta} = (e+p)u_{\alpha}u_{\beta} + pg_{\alpha\beta} +
\frac{1}{\mu_{0}} \left( F_{\alpha \mu} F^{\mu}_{\beta} - \frac{1}{4}
F_{\mu\nu} F^{\mu\nu} \mathrm{g}_{\alpha\beta} \right) \, .
\label{emt}
\ee Here $F_{\alpha\mu}$ is the antisymmetric Faraday tensor defined
as $F_{\alpha\mu} = \partial_{\alpha} A_{\mu} - \partial_{\mu}
A_{\alpha}$, with $A_{\mu}$ denoting the electromagnetic
four-potential $A_{\mu} = (A_{t}, 0 , 0 , A_{\phi})$. The total energy
density of the system is $e$, the pressure is denoted by $p$,
$u_{\alpha}$ is the fluid 4-velocity, and the metric tensor is
$g_{\alpha\beta}$. The first term in Eq.~\eqref{emt} represents the
isotropic (ideal) matter contribution to the energy momentum-tensor,
while the second term is the anisotropic electromagnetic field
contribution.

The metric tensor in axisymmetric spherical-like coordinates $(r,
\theta, \phi)$ can be read of from the line element
\begin{align}
ds^{2} = &-N^{2} dt^{2} + \Psi^{2} r^{2} \sin^{2}\theta (d\phi -
N^{\phi}dt)^{2} \nonumber \\ &+ \lambda^{2}(dr^{2} + r^{2}d\theta^{2}) \, ,
\label{line}
\end{align}
where $N$, $N^{\phi}$, $\Psi$ and $\lambda$ are functions of the
coordinates $(r, \theta)$ \cite{Bonazzola:1993zz}.  As in
Ref.~\cite{Bonazzola:1993zz}, the equation of motion for a star
endowed with magnetic fields reads \be H \left(r, \theta \right) + \nu
\left(r, \theta \right) + M \left(r, \theta \right) = {\rm const},
\label{equationofmotion}
\ee where $H(r,\theta)$ is the heat function defined in terms of the
baryon number density $n$, 
\be H =
\int^{n}_{0}\frac{1}{e(n_{1})+p(n_{1})}\frac{d P}{dn}(n_{1})dn_{1} \, .
\label{heat}
\ee

The quantity $\nu(r, \theta)$ in Eq.~\eqref{equationofmotion} is
defined as $\nu=\ln N$, and the magnetic potential $M(r,\theta)$ is
given by \be M \left(r, \theta \right) = M \left( A_{\phi} \left(r,
\theta \right) \right) \equiv - \int^{0}_{A_{\phi}\left(r, \theta
  \right)} f \left( x \right) dx\, , \ee where $f(x)$ denotes the
current function. Magnetic stellar models are obtained by assuming a
constant value, $f_{0}$, for the latter \cite{Franzon:2016iai}.
According to Ref.~\cite{Bocquet:1995je}, other choices for $f(x)$ are
possible, but the general conclusions as presented in this work remain the same. The constant current function is a standard way to generate self-consistently a dipolar magnetic field
throughout the star.

\begin{figure}[t!]
\centering
  \vspace{0.8cm}
  \includegraphics[width=1.5\textwidth,angle=0,scale=0.48]{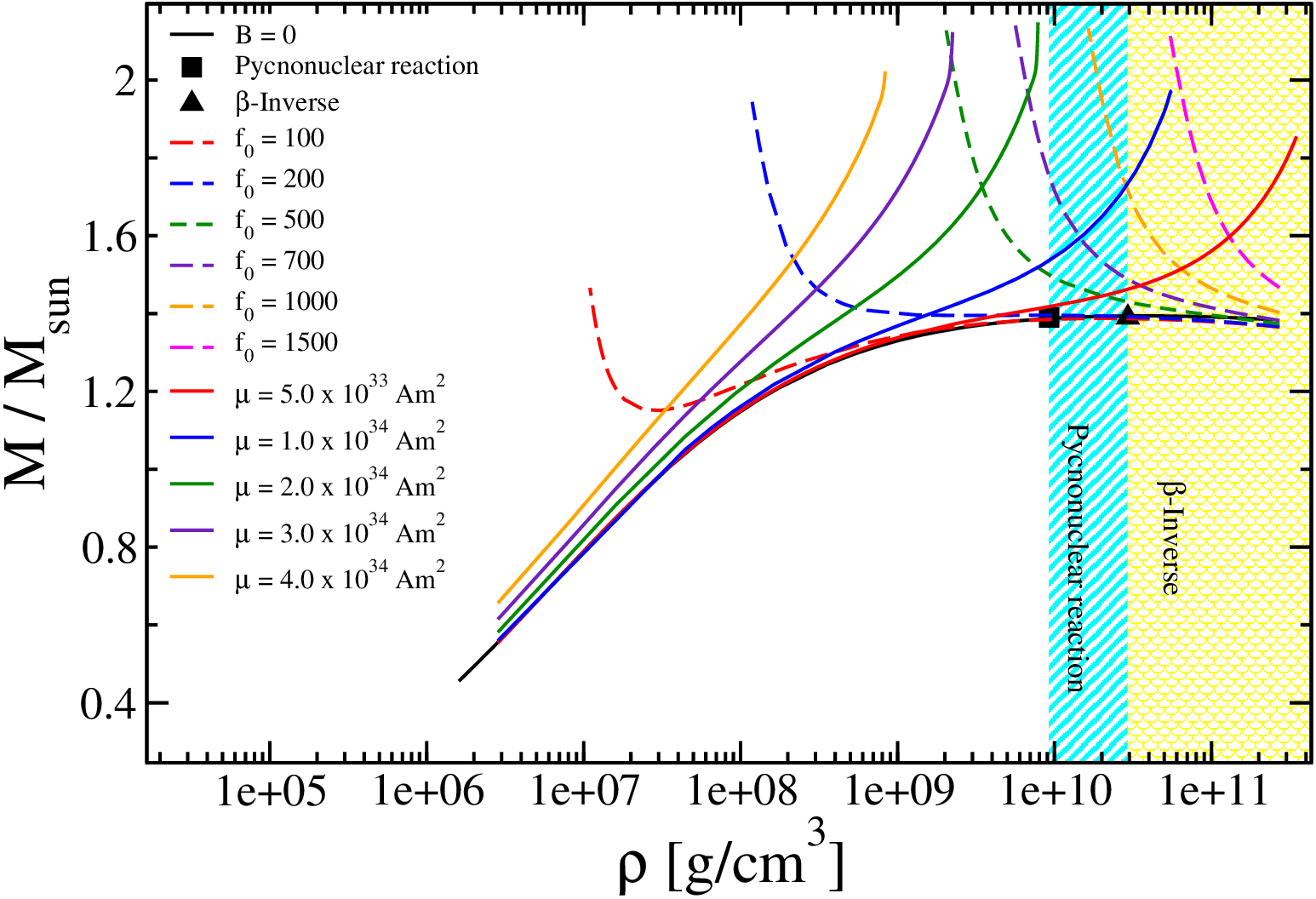}
\caption{(Color online) Gravitational mass as a function of central
  mass density for magnetized white dwarfs, for different values for
  the current function, $f_{0}$, and magnetic dipole moment,
  $\mu$. Stars located in the colored areas are subject to
  pycnonuclear reactions and inverse $\beta$-decay. The threshold of
  these reactions are shown in Table \ref{t1}. The solid square and
  triangle mark the densities at which pycnonuclear and inverse
  $\beta$-decay reactions set in, respectively. %Gravitational
%  instability sets in at the density marked with a solid dot.
}
\label{MRHO}
\end{figure}

\section{Results}\label{sec:results}

In this section, we discuss the effects of strong magnetic fields on
the global properties of stationary white dwarfs taking into account
instabilities due to inverse $\beta$-decay and pycnonuclear fusion
reactions in their cores. In addition, we make use of an equation of
state for white dwarf matter that accounts for electron-ion
interactions and is computed for the latest experimental atomic mass
data.  The instabilities related to the microphysics are fundamental
since they put constraints on the equilibrium configurations and also
limit the maximum magnetic fields which these stars can have
\cite{chamel_stability_2013}. In addition to the magnetic profiles,
which have already been computed in Ref.~\cite{Franzon:2015gda}, we
also compute stellar models at constant magnetic dipole moments
$\mu$. In Ref.~\cite{Franzon:2015gda}, a simple Fermi gas model was
used to model white dwarfs, and the microphysical issues were not
addressed. In our study, the maximum white dwarf mass for
non-magnetized stars is smaller than the one considered in
Ref.~\cite{Franzon:2015gda}, since the lattice contribution softens
the EOS.
\begin{figure}[t!]
\centering
  \vspace{0.8cm}
  \includegraphics[width=1.5\textwidth,angle=0,scale=0.48]{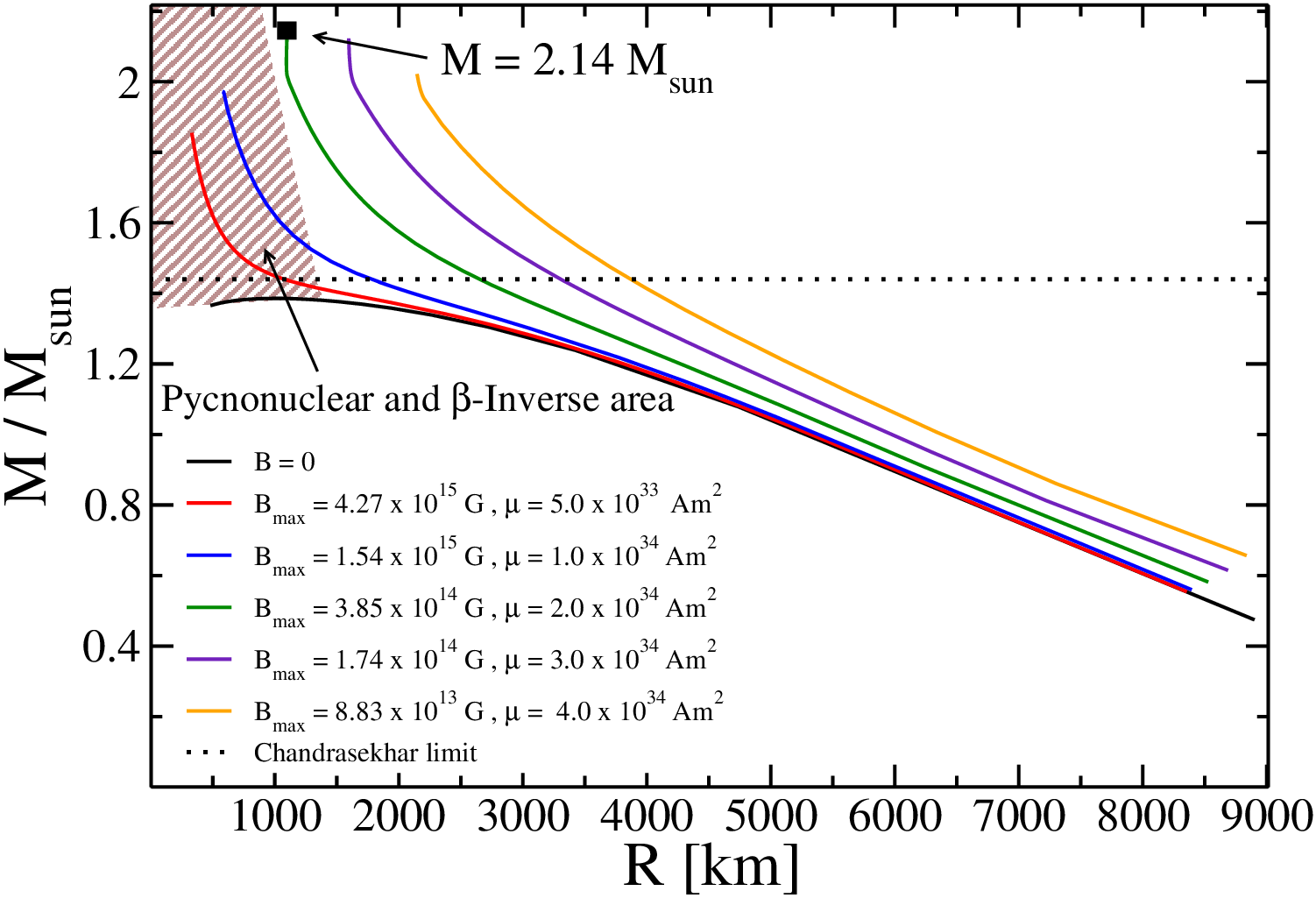}
\caption{(Color online) Mass-radius relationship of magnetized white
  dwarfs for different magnetic dipole moments, $\mu$. The black line
  represents the mass-radius relationship of non-magnetic white
  dwarfs.  The horizontal line represents the Chandrasekhar mass limit
  for spherical stars. Also shown are the values of the magnetic field
  $B_{\rm max}$ (together with the corresponding magnetic dipole
  moment $\mu$) at the centers of the respective maximum mass stars
  (end points of each curve with fixed $\mu$). White dwarfs located in
  the colored (upper left) corner are subject to pycnonuclear fusion
  ($\tau_{\textrm{pyc}}=10$~Gyrs) or inverse $\beta$-decay reactions.}
\label{mr_magmom}
\end{figure}

In Fig.~\ref{MRHO}, we show the gravitational mass versus central
density of white dwarf sequences computed for different (fixed) magnetic
dipole moments, $\mu$, and current functions, $f_{0}$. The magnetic
dipole moment is defined as (see Ref.~\cite{Bonazzola:1993zz})
\begin{equation}
\frac{2\mu \cos\theta}{r^{3}} = B{(r)}\mid_{r\rightarrow \infty} \, ,
\label{mm}
\end{equation}
which is the radial (orthonormal) component of the magnetic field of a
magnetic dipole seen by an observer at infinity. As can be seen from
Fig.~\ref{MRHO}, a larger magnetic moment $\mu$ leads first to an
increase in the white dwarf maximum mass. However, if we increase
$\mu$ further, the maximum mass begins to drop. This is due to the
fact that the stellar radius becomes larger (see also
Fig.~\ref{mr_magmom}), which reduces the magnetic field (see
Eq.~\ref{mm}). As a consequence, the Lorentz force becomes smaller,
rendering the maximum mass configurations less massive.

As can be seen in Fig.\ \ref{MRHO}, the masses of magnetized white
dwarfs increase monotonically with central density. The behavior is
very different if the value of the current function is kept constant,
in which case non-monotonic (in some cases even multivalued)
mass-density relationships are obtained.  The cross-hatched area in
Fig.~\ref{MRHO} shows the density regime where pycnonuclear fusion
reactions become possible. The position of the white dwarf with just
the right threshold density ($9.25\times 10^9$ g/cm$^3$) for this
reaction to occur is marked with a solid black square in
Fig.\ \ref{MRHO}.  The pycnonuclear reaction time at that density is
10 Gyrs. For a central white dwarf density of $1.59\times 10^{10}$
g/cm$^3$ the fusion reaction time decreases to 0.1 Myrs (see Fig. \ref{RhoxM_lam}). 
%The narrow green-hatched band adjacent to the right of the pycnonuclear instability regime shows stars subject to a general
%relativistic instability (labeled GR Instability). \ot{What kind of
%  instability is this? Should we add a sentence here?} \bf{temos que tirar este paragrafo de GR instabildade}
White dwarfs subject to inverse $\beta$-decay reactions in their cores
are located in the yellow area (marked ``$\beta$-inverse'') of
Fig. \ref{MRHO}.  The most massive stable white dwarf which is not
subject to microscopic instability reactions in its core (end point of
the curve with $\mu = 2 \times10^{34}$ Am$^{2}$), has a mass of $\sim
2.14 \, M_{\odot}$ and a radius (see Fig.\ \ref{mr_magmom}) of $\sim
1096$ km.  Finally, we note that the condition $dM/d\rho_{c}>0$ for
stability against radial oscillations is fulfilled for all white dwarf
sequences for which the magnetic dipole moment is kept fixed, as can
be seen in Fig.~\ref{MRHO}.  An overview of the density thresholds
discussed just above is provided  in Table \ref{t1} for white dwarfs with
different magnetic field values and magnetic dipole moments.
\begin{table}[h]
  \centering
  \caption{Thresholds of inverse $\beta$-reactions and pycnonuclear
    fusion reactions (pycnonuclear reaction time of 10 Gyrs) in carbon
    white dwarfs for different magnetic fields, $B$, and magnetic
    dipole moments, $\mu$.}
\label{t1}
\begin{tabular}{cccccc}
\hline
\hline
$^{12}_{6}$C &$\mu$ (Am$^2$) & $B_{\textrm{max}}
$ (G)&$\rho_{\textrm{pyc}}$(g/cm$^3$)&$\rho_{\beta}$(g/cm$^3$)\\
\hline
&$5.0\times 10^{33}$&$4.27\times 10^{15}$ &$9.26\times 10^9$ &$4.00\times 10^{10}$\\
&$1.0\times 10^{34}$&$1.54\times 10^{15}$ &$9.21\times 10^9$ &$4.07\times 10^{10}$\\
&$2.0\times 10^{34}$&$3.85\times 10^{14}$ &$9.24\times 10^9$ &$4.10\times 10^{10}$\\
&$3.0\times 10^{34}$&$1.74\times 10^{14}$ &$9.25\times 10^9$ &$4.10\times 10^{10}$\\
&$4.0\times 10^{34}$&$8.83\times 10^{13}$ &$9.25\times 10^9$ &$4.10\times 10^{10}$\\
\hline
\end{tabular}
\end{table}

\begin{figure*}[t!]
\centering
\vspace{0.1cm}\includegraphics[width=1.6\textwidth,angle=0,scale=0.48]{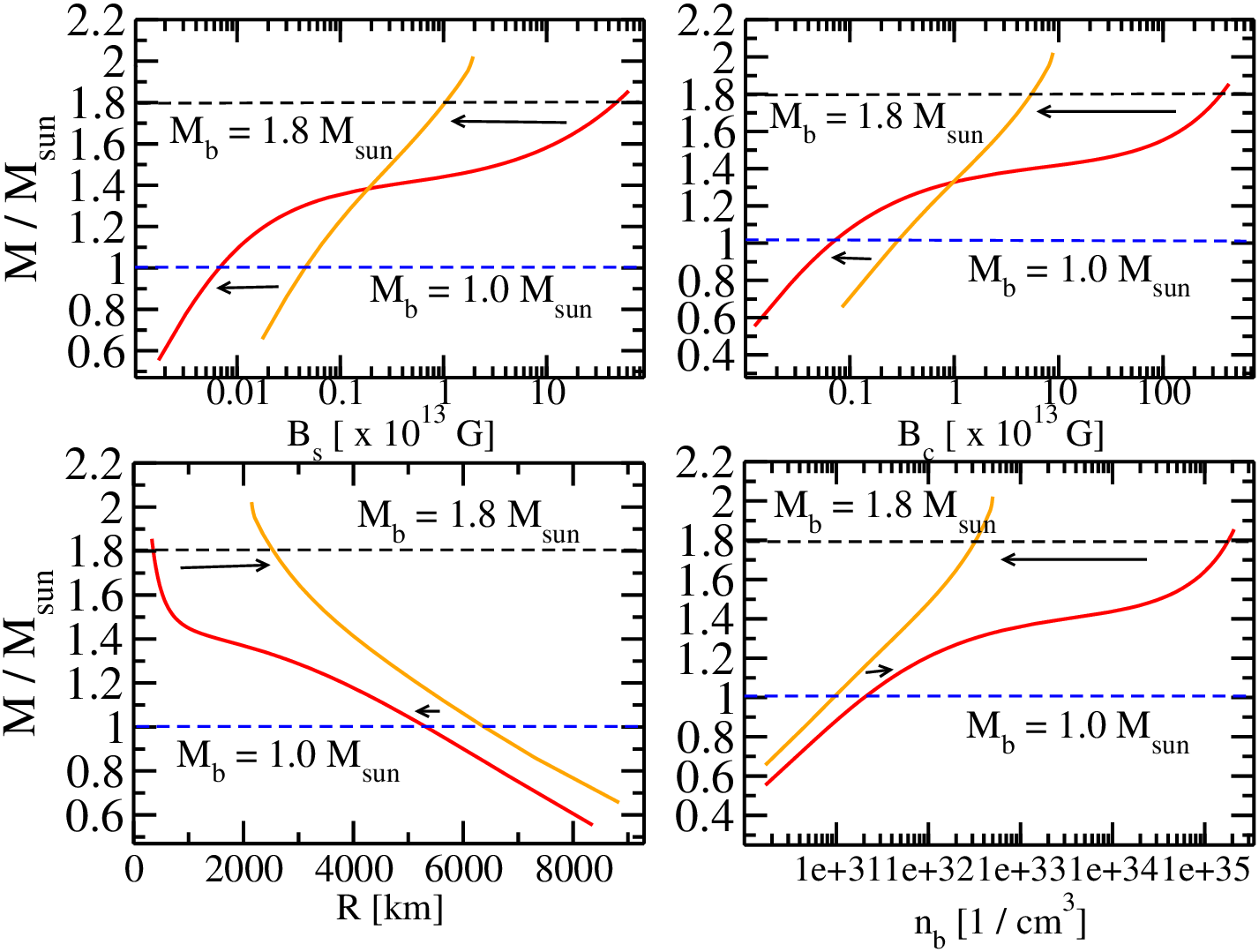}
\caption{(Color online) Global properties of magnetized white dwarfs
  for two different (sample) magnetic dipole moments,
  $\mu=0.5\times10^{34}$ Am$^{2}$ (red line) and
  $\mu=4.0\times10^{34}$ Am$^{2}$ (orange line). $M$ denotes the
  gravitational mass, $B_S$ the magnetic field at the surface, $B_c$
  the magnetic field at the center, $R$ the equatorial radius, and
  $n_b$ the baryon number density.  The horizontal lines represent
  white dwarfs with fixed baryon masses of $M_{B}=1.00\, M_{\odot}$
  (bottom), and $M_{B}=1.80\,M_{\odot}$ (top).  The arrows indicate
  the paths of these white dwarfs in case of a magnetic field
  reduction (see text for details).}
\label{globalproperties1}
\end{figure*}

The mass-radius relationship of magnetized white dwarfs for different
magnetic dipole moments $\mu$ is shown in Fig.~\ref{mr_magmom}. One
sees that increasing values of $\mu$ lead to white dwarfs with larger
radii, because of the added magnetic field energy. The strength of the
magnetic field can be inferred from Fig.~\ref{globalproperties1},
which shows the gravitational mass as a function of surface ($B_{s}$)
and central ($B_{c}$) magnetic fields, the circumferential equatorial
radius ($R_{circ}$), and the baryon number density ($n_{b}$), for two
sample magnetic dipole moments of $\mu=0.5\times10^{34}$~ Am$^{2}$
(red line) and $\mu=4.0\times10^{34}$~ Am$^{2}$ (orange line).

In Fig.\ \ref{globalproperties1} (top panels), one sees that the
curves with $\mu=0.5\times10^{34}~\textrm{Am}^{2}$ and
$\mu=4.0\times10^{34}~\textrm{Am}^2$ cross each other. This is due to
the fact that the magnetic field scales as $\sim \mu / r^{3}$, with
$r$ being the stellar radius (see
Eq.~\eqref{mm}). The locations of stars with fixed baryon masses of
$M_{B}=1.00\, M_{\odot}$ and $M_{B}=1.80\, M_{\odot}$ are shown in
Fig.~\ref{globalproperties1} by dashed horizontal lines. According to
Eq.~\eqref{mm}, the magnetic field is determined by the size of the
star along the curves with $\mu={\rm const}$.  However, along the
lines with fixed baryon masses, the strength of the magnetic field is
a combination of the magnetic dipole moment $\mu$ and the stellar
radius $r$.

Next, we discuss the behavior of the magnetic dipole moments of white
dwarfs whose magnetic fields are weakening. From the $M$ versus $B_s$
and $M$ versus $B_c$ relationships shown in
Fig.~\ref{globalproperties1} (top panels), one sees that two different
scenarios are possible, depending on the star mass and on the magnetic field strengths of white dwarfs. If located above the crossing
point of the $\mu=0.5\times10^{34}~\textrm{Am}^2$ (red line) and
$\mu=4.0\times10^{34}~\textrm{Am}^2$ (orange line) curve, white dwarfs
with weakening magnetic fields would be evolving from right to left
in the two upper panels of Fig.~\ref{globalproperties1}, as shown
(back arrow) for a white dwarf with a constant baryon number of
$M_{B}=1.80\,M_{\odot}$.  The magnetic dipole moment of such white
dwarfs would increase, from $\mu=0.5\times10^{34}$ Am$^{2}$ to
$\mu=4.0\times10^{34}~\textrm{Am}^2$ for the sample star shown in
Fig.~\ref{globalproperties1}.  This is accompanied by an increase in
the stellar radius (see $M$ versus $R$ diagram) and a decrease in the
central baryon density (see $M$ versus $n_{b}$ diagram). The situation
is reversed for white dwarfs located below the the crossing. For such
white dwarfs, a reduction of the magnetic field is accompanied by a
decrease of the magnetic dipole moment, as shown in
Fig.~\ref{globalproperties1} for a sample white dwarf with a constant
baryon mass of $M_{B}=1.00\,M_{\odot}$ (black arrows).  In this case,
white dwarfs become smaller and therefore more dense at the center
(see $M$ versus $R$ and $M$ versus $n_{b}$ diagrams shown in
Fig. \ref{globalproperties1}).

As discussed just above (Fig.~\ref{globalproperties1}), the equatorial
radii of white dwarfs located above the crossing point increase as
their magnetic fields are getting smaller.  The increases in radius
(at a fixed baryon mass) is due to the Lorentz force. However, the
stellar magnetic field scales as $\mu / r^{3}$. This means that for a
star with a mass of $M_{B}=1.80\,M_{\odot}$, the increase in the
magnetic dipole moment, $\mu$, is canceled by the increase in the
radius, reducing the magnetic field. This is the opposite of what is expected for stars with lower
masses. For example, a star with $M_{B}=1.00\,M_{\odot}$ decreases its
magnetic dipole moment and its radius. However, in this case, the
decrease in the radius is not enough to cancel the reduction in
$\mu$. The net result is a decrease of the magnetic field. This can be
understood by looking at the variation in the circular equatorial
radius of the stars with $M_{B}=1.80\,M_{\odot}$ and
$M_{B}=1.00\,M_{\odot}$. For the latter, the change in radius is much
smaller than the radial change for the $M_{B}=1.80\,M_{\odot}$ star,
for a change in the magnetic dipole moment of
$|\Delta\mu|=3.5\times10^{34}$ Am$^{2}$.

\begin{figure*}[t!]
\centering
\vspace{0.1cm}\includegraphics[width=1.6\textwidth,angle=0,scale=0.48]{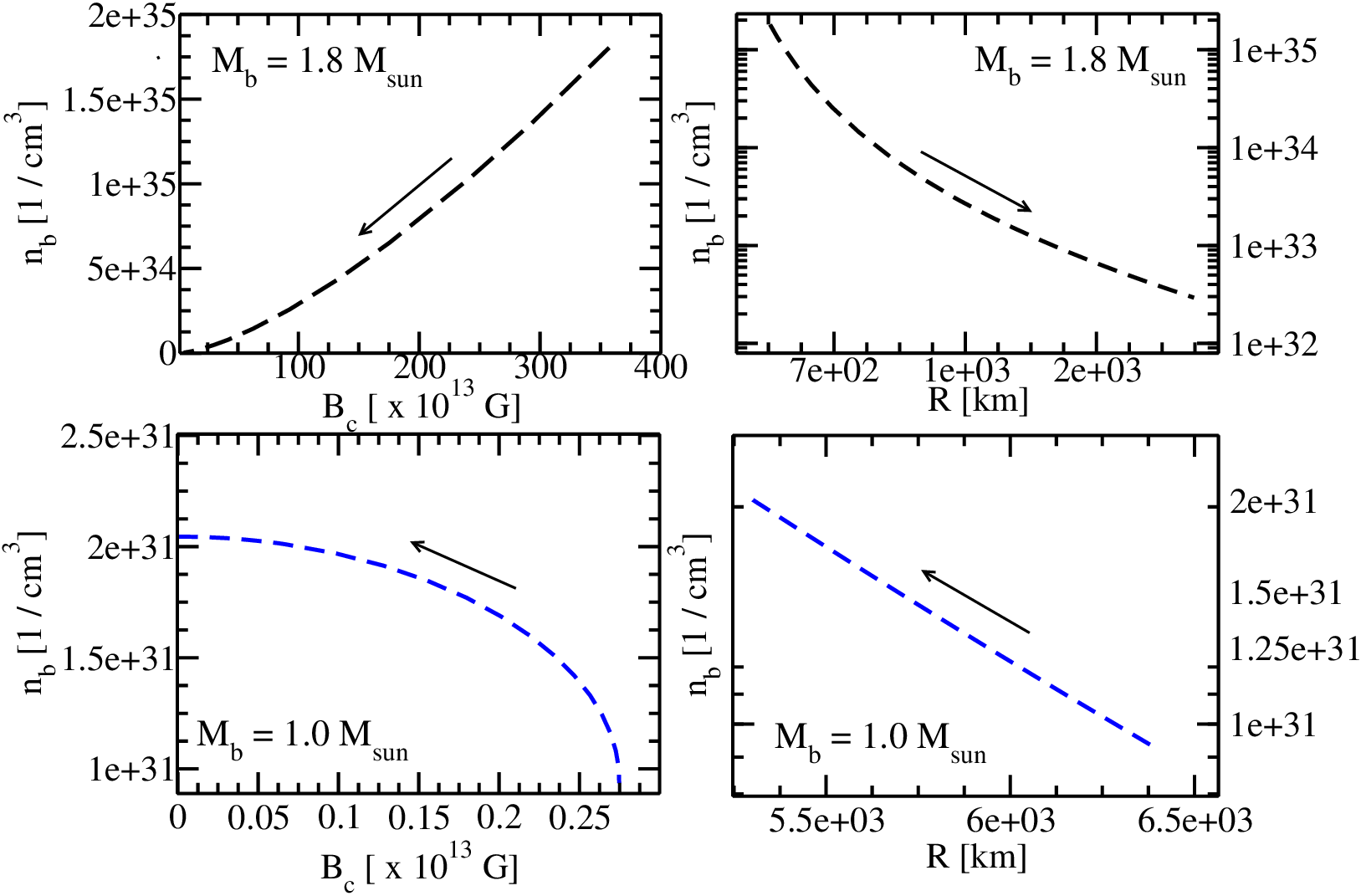}
\caption{Central baryon number density, $n_b$, as a function of
  central magnetic field strength, $B_c$, and equatorial radius, $R$,
  of magnetized white dwarfs with fixed baryon masses of
  $M_{B}=1.00\,M_{\odot}$ and $M_{B}=1.80\,M_{\odot}$. The arrows
  refer to changes in $n_b$ and $R$ for weakening magnetic fields.}
\label{globalproperties2}
\end{figure*}

In Fig.~\ref{globalproperties2}, we show the global properties of two
white dwarfs with fixed baryon masses of $M_{B}=1.00\,M_{\odot}$ and
$M_{B}=1.80\,M_{\odot}$. The top panels show the central baryon
density as a function of the central magnetic field (top-left left
panel) and the circular equatorial radius (top-right panel) for a
white dwarf with $M_{B}=1.80\,M_{\odot}$. For such stars, as the
magnetic field decreases, the central baryon density becomes smaller
due to the fact that the radius is increasing. On the other hand, for
lighter white dwarfs, with a mass of $M_{B}=1.00\,M_{\odot}$, the
central baryon number density increases as the magnetic field
decreases, since the stellar radius is getting smaller.

\section{Summary}\label{sec:summary}

In this work, we presented axisymmetric and stationary models of
magnetized white dwarfs obtained by solving the Einstein-Maxwell
equations self-consistently and taking into stability considerations
related to neutronization due to electron capture reactions as well as
pycnonuclear fusion reactions among carbon nuclei in the cores of
white dwarfs.

 We investigated also  the influence of magnetic fields on the structure of white dwarfs. This is an important problem, since super-massive magnetized WD's, whose existence is partially supported by magnetic forces, could simplify the
explanation of observed ultra-luminous explosions of supernovae Type Ia. The Lorentz force induced by strong magnetic
fields breaks the spherical symmetry of stars and increases their masses, since the force acts in the radial outward direction
against the inwardly directed gravitational pull.
 
In this paper, we make use of an equation of state for a degenerate
electron gas with electron-ion interactions (body-centered-cubic lattice
structure) to describe the matter inside of white dwarfs. We have
shown that the equation of state becomes softer if nuclear lattice
contributions are included in addition to the electron pressure. This
is due to the fact that the repulsive force between electrons is
smaller in the presence of an ionic lattice, causing a softening of
the equation of state (see Fig.~\ref{eos_carbono}). We note that the
density thresholds for pycnonuclear fusion reactions and inverse
$\beta$-reactions are reduced when magnetic fields are present in the
stellar interior, as can be seen in Table \ref{t1}.

We have shown that the masses of white dwarfs increase up to $M=2.14
\, M_{\odot}$ (with a corresponding magnetic dipole moment of
$\mu=2.0\times 10^{34}$ Am$^2$ (see, e.g., Fig.~\ref{MRHO}) if
microphysical instabilities are considered. This star has an
equatorial radius of $\sim 1100$ km with magnetic fields of
$B_{c}=3.85\times10^{14}$ G and $B_{s}=7.21\times10^{13}$ G at the
center and at the stellar surface, respectively. For this white dwarf,
the ratio between the magnetic pressures and the matter pressure at
the center is 0.789. Although the surface magnetic fields obtained
here are higher than the observed ones for white dwarfs, these figures
provide an idea of the maximum possible magnetic field strength that
can be reached inside of these objects, and may also be used to assess
the effects of strong magnetic fields on both the microphysics and the
global structure of magnetized white stars.

The maximum magnetic field found in this work is an order of magnitude
smaller than that of Ref.~\cite{Franzon:2015gda}. This is because we
modeled the stellar interior with a more realistic equation of state
than just a simple electron gas, and we considered the density
threshold for pycnonuclear fusion reactions for a 10 Gyrs fusion
reaction time scale, which restricts the central density of white
dwarfs to $\sim 9.25\times 10^9$ g/cm$^3$ (see Table \ref{t1}),
limiting the stellar masses and, therefore, their radii, which for
very massive and magnetized white dwarfs cannot be smaller than R$\sim
1100$ km. However, it is important to mention that the pycnonuclear
reaction time scales are somewhat uncertain. In our case, for example,
we have a factor of uncertainty of approximately 3.5 in the
calculation of the astrophysical S-factor (see
Refs. \cite{gasques_nuclear_2005,yakovlev_fusion_2006}).

Our results show that the surface magnetic field, $B_s$, is about one
  order of magnitude smaller than the magnetic field reached at the
  stellar center, $B_c$. If the magnetic field weakens for massive
  white dwarfs, we found that the magnetic dipole moments of such
  stars may increase (Fig.\ \ref{globalproperties1}), which is due to
  the fact that, for a fixed baryon mass, the magnetic field is
  determined by the interplay between the magnetic dipole moment and
  the stellar radius.  The situation is reversed for less massive
  white dwarfs, for which smaller the magnetic fields imply smaller
  stellar magnetic dipole moments.
The radii of massive (light) white dwarfs are found to increase
(decrease) for decreasing central magnetic fields
(Fig.\ \ref{globalproperties2}).  This opens up the possibility that
massive white dwarfs, with central magnetic fields greater than
$B\sim 10^{13}$ G, increase their magnetic fields through continued
compression.  This phenomenology differs from previous studies carried
out for magnetic fields less than $\sim 10^{13}$ G
\cite{suh2000mass,ostriker1968rapidly1}, where an increase of the
central magnetic field was found to make stars less dense and
therefore bigger in size.

We note that stellar configurations which contain only poloidal
magnetic fields (no toroidal component) are unstable (see, e.g.,
\cite{armaza2015magnetic, mitchell2015instability,
  braithwaite2006stability}).  Moreover, according to
Ref.~\cite{goldreich_magnetic_1992}, many different mechanisms can
affect the magnetic fields and their distributions inside of white
dwarfs. In this work, in the framework of a fully general relativistic
treatment, we model the properties of magnetized white dwarfs with
purely poloidal magnetic field components. Although this is not the
most general magnetic field profile, and a dynamical stability of
these stars still needs to be addressed, magnetic fields considerably
increase the masses of white dwarfs, even when microphysical
instabilities are considered. As a consequence, such white dwarfs
ought to be considered as possible candidates of super-Chandrasekhar
white dwarfs, thereby contributing to our understanding of
superluminous type-Ia supernovae.

Lastly, we note that for a typical magnetic field value of $\sim
10^{14}$ G and a density of $\sim10^{9}$ g/cm$^3$, we obtain an Alfven
velocity of $v=10^{9}$ cm/s, which, for a white dwarf with a typical
radius of $R=1500$ km, leads to an Alfven crossing time of $\sim 0.1$
s \cite{durisen1973viscous, yakovlev1980thermal, cumming2002magnetic}.
This is close to the hydrostatic equilibration time of white
dwarfs. As a consequence, although magnetized white dwarfs seem to be
short-lived stars, they might still be supported by magnetic
fields. Our results represent magnetostatic equilibrium
conditions. The stability analysis of such systems is beyond the scope
of this study, which constitutes a first step toward a more complete
discussion of the possible existence of super-Chandrasekhar white
dwarfs. Studies which address issues such as the role of different
(poloidal and toroidal) magnetic field configurations, stellar
rotation, and different compositions of the stellar cores will be
presented in a series of forthcoming papers.
 
%%%%%%%%%%%%%%%%%%%%%%%%%%%%%%
\section{Acknowledgments}
We acknowledge financial support from the Brazilian agencies CAPES,
CNPq, and we would like to thank FAPESP for financial support under
the thematic project 13/26258-4 B.\ Franzon acknowledges support from CNPq/Brazil, DAAD and HGS-HIRe for FAIR.  S.\ Schramm acknowledges support from the HIC for FAIR LOEWE program.  F.\ Weber is supported by the National
Science Foundation (USA) under Grant PHY-1411708.

\bibliography{Ref}
\end{document}